# Bending mode and thermal expansion of graphene

V.N. Bondarev[1,†], V.M. Adamyan[1,‡], V.V. Zavalniuk[1,2,*]

[1] *Department of Theoretical Physics,
Odessa I.I. Mechnikov National University, 2 Dvoryanska St., Odessa 65026, Ukraine*
[2] *Department of Fundamental Sciences,
Odessa Military Academy, 10 Fontanska Road, Odessa 65009, Ukraine*

## Abstract

Proceeding from the model of a two-dimensional elastic continuum, we describe the characteristic features of thermal expansion of graphene using an approach that goes beyond the quasi-harmonic approximation. The negative value of the thermal expansion coefficient of graphene at low temperatures and its sign reversal at $T \approx 1000$ K are established. It is shown that the bending vibrational mode plays a decisive role in peculiarities of the thermal contraction/expansion of graphene and that the contribution of this mode to the thermal expansion coefficient does not depend on the sample size, due to the "sound" nature of the bending mode long-wave dispersion. The obtained results allow giving a quantitative description of the known data of numerical experiments on the thermal expansion of graphene in the entire interval of the MD simulations.

## I. INTRODUCTION

Although most solids expand when heated (see, for example, [1]), the volumetric thermal expansion coefficient (TEC) of some bulk crystals (e.g., Si, Ge, β-quartz) can be negative [2]. For example, the TEC of graphite in the hexagonal plane is negative for $T \lesssim 600$ K (see [3, 4] and references therein). The investigation of graphene, which is a single graphite layer, showed that this 2D crystal exhibits no less surprising behavior than graphite (see, for example, [5, 6]). Here, under graphene TEC (GTEC) we understand the thermal expansion of the graphene sample projection to the reference $x - y$ plane where it lies presumably at $T = 0$. The temperature dependence of GTEC has been studied in many papers beginning with [3], where the calculations by the density functional method showed that TEC was negative at least up to $\approx 2300$ K. Later, it was reported in [7] that the atomistic Monte Carlo simulations indicate the GTEC crossover from minus to plus at $T \approx 900$ K. The experimental detection of the analogous GTEC change of sign but in the interval 350 – 400 K was announced in [8, 9]. However, the values of GTEC estimated in [10] on the base of temperature-dependent Raman spectroscopy are unlike [8, 9] all negative between 200 and 400 K. So the sign of GTEC at low temperatures is almost no doubt negative [10,11] though the crossover at some temperature is still debatable (see, for example, [12]).

The numerical similarity between the "low-temperature" TEC values for graphene and graphite (in the hexagonal plane) noted in [3] led to the conclusion that the effect of thermal contraction is due to the presence of the bending mode in both allotropes. Until recently the dispersion of the bending mode of graphene was considered as quadratic at $q \to 0$, by analogy with flexural vibrations of elastic plates [13]. Unfortunately, the model of the bending mode with $\sim q^2$ dispersion inevitably leads to the fact that the mean-square fluctuations of the out-of-plane displacements of a quasi-2D crystal with a characteristic linear size $R$ are "catastrophic" in the limit $R\to\infty$, i.e. they diverge as $\sim R^{2\zeta}$ ($\zeta = 1$ in a simple membrane model [14]; more sophisticated calculations give $\zeta \approx 0.6$ [15]). Moreover, an attempt to construct a model of the thermal expansion (contraction) of graphene under the assumption of the quadratic dispersion of the bending (out-of-plane) mode results in a divergent (due to small $q$) negative TEC (the discussion of this issue, see, for example, in [4,16-18]). On the other hand, the results of numerical simulations of the graphene mean-square out-of-plane displacements [19] do not show any signs of the mentioned "catastrophic" in $R$ divergences.

[*] vzavalnyuk@onu.edu.ua (corresponding author)
[†] bondvic@onu.edu.ua
[‡] vadamyan@onu.edu.ua

In this paper we deduced an appropriate explicit expression for GTEC as function of temperature and thus eliminate the marked vagueness. To achieve this goal we modeled the free standing graphene as elastic continuum [20] retaining in the corresponding energy operator only the contributions of the quadratic, cubic and some biquadratic terms in derivatives of components of the displacement vector. Assuming that the graphene sample was spontaneously subjected to static deformation, we considered the components of the static strain tensor as thermodynamic variables with respect to which the free energy of the sample must have a minimum. The found in this way equilibrium value of static part of the strain tensor yields the desired expression for GTEC. It indeed turns out to be negative at low and room temperatures, but changes sign to positive at $T \approx 1100$ K if one substitutes into the obtained formula for GTEC an appropriate second and third order elastic moduli of graphene and take into account the quasi-harmonic terms as well.

Note that, as has been proved in [21], the dispersion of bending mode in graphene is actually acoustic due to the specific fluctuation interaction between in-plane and out-of-plane vibrations conditioned by non-linear components in the graphene strain tensor. The conclusions of paper [21] were confirmed in Refs. 22-25, devoted to the study of mechanical and vibrational characteristics of graphene by means of computer simulations. A direct consequence of this fact is that GTEC does not depend on the size of graphene sample.[1] This should be taken into account when interpreting numerical results for the thermal properties of graphene.

Most of theoretical models of TEC of solid do not go beyond the so-called quasi-harmonic approximation (see, [1, 20, 26]), which is simply the harmonic approximation for any values of the lattice constant with assumed dependence of the latter on the pressure. With this approach, however, the effect of anharmonic in strain tensor terms (first of all, cubic) in the crystal Hamiltonian on the thermal expansion is lost (the role of third-order terms in thermal expansion of crystals is briefly discussed in book [27]). Meanwhile, as we will see below on the example of graphene, such terms make a considerable contribution to the thermal expansion of quasi-2D crystals, and at low temperatures their role in this phenomenon is decisive.[2]

This paper is organized in the following way. In Section II using the model of elastic continuum we find an expression for the free energy due to both the vibrational modes and the static (resulting in thermal expansion) displacements of the graphene lattice, with explicit consideration of the third-order contributions over the strain tensor of a quasi-2D crystal. In Section III, these results are applied to the calculation of GTEC, and it is shown that at temperatures up to about 1000 K, graphene should experience the thermal contraction, due to the presence in GTEC of a *strictly negative* contribution[3] arising beyond the quasi-harmonic approximation. With increasing temperature, the "quasi-harmonic" contribution to GTEC becomes significant, providing a tendency to reach a certain positive value of GTEC at high temperatures. The constructed theory is applied to the quantitative description of numerical "experiments" on the thermal expansion of graphene. Finally, in Section IV, we sum up the results obtained and outline directions for further research.

---

[1] In the mentioned works [17, 18], a possibility of linear dispersion at small $q$ for the graphene bending mode was considered. So, the result of Ref. 17 was based on using the "membrane", $\sim \kappa q^2$, spectrum in the initial step, with further account of the third and fourth order terms in the in-plane and out-of-plane displacements. Therefore, the "velocity" of the bending mode obtained in [17] was substantially determined by the so-called bending rigidity $\kappa$. The use of a renormalized sound-like spectrum of the graphene bending mode led in [17] to a convergent negative value for the low-temperature GTEC. In Ref. 18, the linear dispersion of the bending mode at $q\to 0$ has been obtained in the framework of a discrete atomistic model, taking into account the anharmonic coupling of third and fourth orders in the "Hamiltonian" of graphene. However, the "velocity" of the bending mode found in [18] turned out to be divergent (although, weakly) with the growth of the linear size $R$ of the graphene crystal. A similar divergence was manifested in the GTEC calculated in the framework of the approach [18].

[2] It was shown in [28] that the addition of anharmonic terms into Hamiltonian allows explaining the existence of a temperature region (up to $\approx 600$ K), in which GTEC turns out to be negative; see also [29], where accounting for anharmonic effects led to negative values of GTEC at $T < 360$ K.

## II. BASIC CONCEPTS: ANHARMONIC EFFECTS AND THERMAL AVERAGES IN QUASI-2D ELASTIC CONTINUUM

In the present paper, we consider graphene as an infinitesimally thin elastic film that lies at equilibrium in the reference plane $x - y$. Each graphene point mass is associated with the 2D radius-vector $\mathbf{r} = (x, y)$ of its equilibrium position. The shift of the point mass to a new position in three-dimensional space at time $t$ due to deformation is determined by the displacement vector $(\mathbf{u}(\mathbf{r},t), w(\mathbf{r},t))$, where $\mathbf{u}(\mathbf{r},t) = (u_x(\mathbf{r},t), u_y(\mathbf{r},t))$ and $w(\mathbf{r},t)$ are its projections onto $x - y$ plane and $z$-axis (in-plane and out-of-plane components), respectively. We define, as usually for continuous media, the velocities $\dot{\mathbf{u}}(\mathbf{r},t)$ and $\dot{w}(\mathbf{r},t)$ of a graphene point mass by material derivatives given by relations

$$\dot{\mathbf{u}}(\mathbf{r},t) = \frac{\partial \mathbf{u}(\mathbf{r},t)}{\partial t} + (\dot{\mathbf{u}}(\mathbf{r},t) \cdot \nabla)\mathbf{u}(\mathbf{r},t), \quad \dot{w}(\mathbf{r},t) = \frac{\partial w(\mathbf{r},t)}{\partial t} + (\dot{\mathbf{u}}(\mathbf{r},t) \cdot \nabla)w(\mathbf{r},t), \quad (1)$$

where $\nabla$ is 2D- gradient. In what follows we use instead of (1) the approximation

$$\dot{\mathbf{u}}(\mathbf{r},t) = \frac{\partial \mathbf{u}(\mathbf{r},t)}{\partial t} + \left(\frac{\partial \mathbf{u}(\mathbf{r},t)}{\partial t} \cdot \nabla\right)\mathbf{u}(\mathbf{r},t), \qquad \dot{w}(\mathbf{r},t) = \frac{\partial w(\mathbf{r},t)}{\partial t} + \left(\frac{\partial \mathbf{u}(\mathbf{r},t)}{\partial t} \cdot \nabla\right)w(\mathbf{r},t).$$

The in-plane components of the graphene strain tensor $\varepsilon_{\alpha\beta}(\mathbf{r},t)$, where each Greek index runs the set $(x, y)$, according to definition [13] are

$$\varepsilon_{\alpha\beta}(\mathbf{r},t) = \frac{1}{2}[\partial_\alpha u_\beta(\mathbf{r},t) + \partial_\beta u_\alpha(\mathbf{r},t) + \partial_\alpha u_\gamma(\mathbf{r},t)\partial_\beta u_\gamma(\mathbf{r},t) + \partial_\alpha w(\mathbf{r},t)\partial_\beta w(\mathbf{r},t)], \quad \partial_\alpha \equiv \frac{\partial}{\partial r_\alpha}.$$

With these notations and definitions we model the "mechanical" part $\boldsymbol{H}\{\mathbf{u},w\}$ of graphene Hamiltonian as the operator functional

$$\boldsymbol{H}\{\mathbf{u},w\} = \int d\mathbf{r} \left\{ \frac{\rho}{2}[\dot{\mathbf{u}}^2(\mathbf{r},t) + \dot{w}^2(\mathbf{r},t)] + \frac{\lambda}{2}\varepsilon_{\alpha\alpha}(\mathbf{r},t)\varepsilon_{\beta\beta}(\mathbf{r},t) + \mu\varepsilon_{\alpha\beta}(\mathbf{r},t)\varepsilon_{\alpha\beta}(\mathbf{r},t) + \frac{\kappa}{2}[\nabla^2 w(\mathbf{r},t)]^2 \right.$$
$$\left. + \frac{C_{111} - C_{112}}{4}\varepsilon_{\alpha\beta}(\mathbf{r},t)\varepsilon_{\alpha\beta}(\mathbf{r},t)\varepsilon_{\gamma\gamma}(\mathbf{r},t) + \frac{3C_{112} - C_{111}}{12}\varepsilon_{\alpha\alpha}(\mathbf{r},t)\varepsilon_{\beta\beta}(\mathbf{r},t)\varepsilon_{\gamma\gamma}(\mathbf{r},t) \right\}, \quad (2)$$

where the summation is assumed over repeated indices, $\rho$ is the two-dimensional mass density of graphene, $\mu > 0$ and $\lambda > -\mu$ are two-dimensional Lamé coefficients (in fact, $\lambda > 0$, see [13]), $C_{111}$, $C_{112}$ are the third order elastic coefficients. Formally, the cubic terms in the strain tensor $\varepsilon_{\alpha\beta}(\mathbf{r},t)$ for the hexagonal symmetry of graphene should stand in the Hamiltonian (2) with three elasticity moduli of third order: $C_{111}$, $C_{112}$, and $C_{222}$ [30]. However, for simplicity, we write them as in the isotropic model, i.e. put $C_{111} = C_{222}$ (according to calculations in [30], for graphene indeed $C_{111} \approx C_{222} < 0$ and $C_{112} \approx C_{111}/3$; see also the experimental work [31]).

---

[3] We emphasize that this negative contribution would be divergent [16] in the limit of long waves for a bending mode with dispersion of $\sim q^2$. And only the appearance of a "sound" dispersion of the bending mode of a quasi-2D crystal [21] eliminates a similar IR divergence of GTEC.

In order to apply the described model of the elastic continuum for identification of the features of the thermal expansion/contraction of graphene, suppose that the graphene elastic film spontaneously undergoes static deformation and represent the in-plane part $\mathbf{u}(\mathbf{r},t)$ of displacement vector as the sum

$$\mathbf{u}(\mathbf{r},t) = \mathbf{u}^{(s)}(\mathbf{r}) + \mathbf{u}^{(v)}(\mathbf{r},t)$$

of the static displacement (expansion/contraction) $\mathbf{u}^{(s)}(\mathbf{r})$ and dynamic (vibrational) $\mathbf{u}^{(v)}(\mathbf{r},t)$ parts. Introducing the linear tensors of static and dynamic strains (below we omit the arguments of the field functions)

$$u^{(s)}_{\alpha\beta} \equiv \frac{1}{2}(\partial_\alpha u^{(s)}_\beta + \partial_\beta u^{(s)}_\alpha), \quad u^{(v)}_{\alpha\beta} \equiv \frac{1}{2}(\partial_\alpha u^{(v)}_\beta + \partial_\beta u^{(v)}_\alpha)$$

we can write the Hamiltonian (2) in the form:

$$\boldsymbol{H}\{\mathbf{u},w\} \equiv \boldsymbol{H}^{(v)}\{\mathbf{u}^{(v)},w\} + \boldsymbol{H}^{(s)}\{\mathbf{u}^{(s)}\} + \boldsymbol{H}^{(s,v)}\{\mathbf{u}^{(s)},\mathbf{u}^{(v)},w\}, \tag{3}$$

where

$$\boldsymbol{H}^{(v)}\{\mathbf{u}^{(v)},w\} = \int d\mathbf{r}\left\{\frac{\rho}{2}\left[\left(\frac{\partial \mathbf{u}^{(v)}}{\partial t}\right)^2 + \left(\frac{\partial w}{\partial t}\right)^2\right] + \frac{\lambda}{2}u^{(v)}_{\alpha\alpha}u^{(v)}_{\beta\beta} + \mu u^{(v)}_{\alpha\beta}u^{(v)}_{\alpha\beta} + \frac{\kappa}{2}(\nabla^2 w)^2\right.$$

$$+ \frac{1}{4}(\nabla w)^2\left(\lambda \partial_\alpha u^{(v)}_\beta \partial_\alpha u^{(v)}_\beta + \frac{3C_{112}-C_{111}}{2}u^{(v)}_{\alpha\alpha}u^{(v)}_{\beta\beta} + \frac{C_{111}-C_{112}}{2}u^{(v)}_{\alpha\beta}u^{(v)}_{\alpha\beta}\right)$$

$$\left. + \frac{1}{2}\partial_\alpha w \partial_\beta w\left(\mu \partial_\alpha u^{(v)}_\gamma \partial_\beta u^{(v)}_\gamma + \frac{C_{111}-C_{112}}{2}\partial_\alpha u^{(v)}_\beta \partial_\gamma u^{(v)}_\gamma + \rho\frac{\partial u^{(v)}_\alpha}{\partial t}\frac{\partial u^{(v)}_\beta}{\partial t}\right)\right\}, \tag{4}$$

$$\boldsymbol{H}^{(s)}\{\mathbf{u}^{(s)}\} = \int d\mathbf{r}\left(\frac{\lambda}{2}u^{(s)}_{\alpha\alpha}u^{(s)}_{\beta\beta} + \mu u^{(s)}_{\alpha\beta}u^{(s)}_{\alpha\beta}\right), \tag{5}$$

$$\boldsymbol{H}^{(s,v)}\{\mathbf{u}^{(s)},\mathbf{u}^{(v)},w\} = \int d\mathbf{r}\left\{\partial_\alpha u^{(s)}_\beta\left(\lambda \partial_\alpha u^{(v)}_\beta u^{(v)}_{\gamma\gamma} + \mu \partial_\alpha u^{(v)}_\gamma \partial_\gamma u^{(v)}_\beta\right)\right.$$

$$+ \frac{1}{2}u^{(s)}_{\alpha\alpha}\left[\lambda\left(\partial_\beta u^{(v)}_\gamma \partial_\beta u^{(v)}_\gamma + (\nabla w)^2\right) + \frac{C_{111}-C_{112}}{2}u^{(v)}_{\beta\gamma}u^{(v)}_{\beta\gamma} + \frac{3C_{112}-C_{111}}{2}u^{(v)}_{\beta\beta}u^{(v)}_{\gamma\gamma}\right]$$

$$\left. + u^{(s)}_{\alpha\beta}\left[\mu\left(\partial_\alpha u^{(v)}_\gamma \partial_\beta u^{(v)}_\gamma + \partial_\gamma u^{(v)}_\alpha \partial_\gamma u^{(v)}_\beta + \partial_\alpha w \partial_\beta w\right) + \frac{C_{111}-C_{112}}{2}u^{(v)}_{\alpha\beta}u^{(v)}_{\gamma\gamma} + \rho\frac{\partial u^{(v)}_\alpha}{\partial t}\frac{\partial u^{(v)}_\beta}{\partial t}\right]\right\}, \tag{6}$$

In (4)-(6) only the terms of first and second orders in powers of static displacements are retained. Besides, we omitted all the terms in the Hamiltonian, which in the approximation considered below either do not affect the free energy, or lead to some temperature variations of the elastic moduli $\lambda$ and $\mu$. Finally, we retained in (4) the terms of the form $\partial_\alpha u^{(v)}_\gamma \partial_\beta u^{(v)}_\gamma \partial_\alpha w \partial_\beta w$, which are responsible for the appearance in graphene of a nonzero bending elastic modulus [21]. We emphasize that the "cross" contributions (6), which describe an interplay between the static and dynamic strains in graphene, are generally absent in the quasi-harmonic approximation and do appear only if at least the third-order terms in strain tensor are explicitly taken into account in the Hamiltonian of a solid (cf. the discussion of similar problem in [27]).

Starting from (4), we must first refine the explicit expression for the bending elastic modulus $B$ obtained in [21]. It was shown there that the acoustic-type dispersion of bending mode in graphene is generated by the fluctuation interaction between in-plane and out-of-plane terms in (4). Using an original adiabatic approximation based on the confirmed *a posteriori* significant difference of sound speeds for in-plane and bending modes we can replace (4) with the effective "harmonic" Hamiltonian

$$H_B^{(v)}\{\mathbf{u}^{(v)},w\} = \int d\mathbf{r} \left[ \frac{\rho}{2}\left(\frac{\partial \mathbf{u}^{(v)}}{\partial t}\right)^2 + \frac{\lambda}{2} u_{\alpha\alpha}^{(v)} u_{\beta\beta}^{(v)} + \mu u_{\alpha\beta}^{(v)} u_{\alpha\beta}^{(v)} + \frac{\rho}{2}\left(\frac{\partial w}{\partial t}\right)^2 + \frac{B}{2}(\nabla w)^2 + \frac{\kappa}{2}(\nabla^2 w)^2 \right] \quad (7)$$

with the effective bending modulus

$$B = \frac{1}{4}\left(3\lambda + 5\mu + \frac{3C_{111}+C_{112}}{4}\right)\langle \partial_\alpha u_\beta^{(v)} \partial_\alpha u_\beta^{(v)} \rangle_{\mathbf{u}} \quad (8)$$

and $\langle ... \rangle_{\mathbf{u}}$ denotes the thermal average for the density operator with the same Hamiltonian (7). Expression (8) differs from that found in paper [21] by the presence of additional term $\lambda + 3\mu$ in parentheses. This difference is caused by the explicit account in (7) of the terms

$$\frac{\rho}{2} \partial_\alpha w \, \partial_\beta w \, \frac{\partial u_\alpha^{(v)}}{\partial t} \frac{\partial u_\beta^{(v)}}{\partial t}.$$

Note that the in-plane and out-of-plane modes are separated in the resulting expression (7).

Moreover, expression (8) can be formally obtained if we consider the bending modulus $B$ as a free thermodynamic parameter and determine its value from the condition of free energy minimum, considering the discrepancy between the Hamiltonians (4) and (7) as a perturbation or using the Bogolyubov-Peierls variational principle.

Accepting for graphene $\rho \approx 7.6\times 10^{-8}$ g/cm$^2$, $\mu \approx 3\lambda \approx 9$ eV/Å$^2$, we get for the longitudinal $s_L$ and transverse $s_T$ speeds of in-plane waves in graphene the values [32] $s_L = \sqrt{(\lambda+2\mu)/\rho} \approx 21$ km/s and $s_T = \sqrt{\mu/\rho} \approx 14$ km/s.

Introducing the average in-plane sound velocity $2s_\parallel^{-2} = s_L^{-2} + s_T^{-2}$, the maximum wave number $k_{max} = \sqrt{4\pi\rho/m}$ (for graphene $m = 2\times 10^{-23}$ g is the $^{12}$C atomic mass), and the "in-plane Debye temperature"

$$\theta_\parallel \equiv \hbar s_\parallel k_{max} = 2\hbar\sqrt{\frac{2\pi\mu(\lambda+2\mu)}{m(\lambda+3\mu)}} = 2670 \text{ K} \quad (9)$$

and using the standard approach of the solid state theory we get

$$B = \frac{\hbar\sqrt{\pi\rho}}{3m^{3/2}s_\parallel}\left(3\lambda + 5\mu + \frac{3C_{111}+C_{112}}{4}\right)\left[1 + 6\left(\frac{T}{\theta_\parallel}\right)^3 \int_0^{\theta_\parallel/T} \frac{\xi^2 d\xi}{e^\xi - 1}\right]. \quad (10)$$

From the "Hamiltonian" (7) it follows the dispersion law of the bending mode:

$$\omega_B^2(q) = s_B^2 q^2 + \frac{\kappa}{\rho} q^4, \qquad s_B = \sqrt{\frac{B}{\rho}}. \tag{11}$$

By (10) and (11) the natural condition of the thermodynamic stability of graphene demands

$$3\lambda + 5\mu + \frac{3C_{111} + C_{112}}{4} > 0. \tag{12}$$

In this regard, the following clarification should be made. In [30], there were presented the results of numerical calculations of the third-order elastic moduli $C_{111}$ and $C_{112}$ for graphene according to which $(3C_{111} + C_{112})/4 \approx -1250$ N/m. These moduli, in their sense (as well as all the quantities appearing in the theory of elasticity [13]), should be attributed to the "acoustic" degrees of freedom. Meanwhile, in the framework of the generally accepted Debye model [26, 27] for crystals (including graphene) which also have optical modes, $C_{111}$ and $C_{112}$ should be understood as some effective quantities. This means that their values in the above expressions may not coincide with those found in [30]. To estimate the effective value of $(3C_{111} + C_{112})/4$, we use the "low-temperature" value of the bending sound velocity in graphene from [22]: $s_B \approx 0.3$ km/s (see also [21]), and as a result we get: $(3C_{111} + C_{112})/4 \approx -780$ N/m. However, it should be mentioned that, in contrast to our theory, numerical simulation [22] was built on a purely classical basis.

The total free energy of quasi-2D graphene with account of both oscillations and the thermal expansion of the lattice can be obtained from the fundamental formula [26]

$$F = -T \ln \mathrm{Tr}\left[\exp\left(-\frac{\boldsymbol{H}_B^{(v)}\{\mathbf{u}^{(v)}, w\} + \boldsymbol{H}^{(s)}\{\mathbf{u}^{(s)}\} + \boldsymbol{H}^{(s,v)}\{\mathbf{u}^{(s)}, \mathbf{u}^{(v)}, w\}}{T}\right)\right], \tag{13}$$

where the purely "dynamic" part $\boldsymbol{H}^{(v)}\{\mathbf{u}^{(v)}, w\}$ from (3) is replaced by the "harmonic" operator $\boldsymbol{H}_B^{(v)}\{\mathbf{u}^{(v)}, w\}$. To find the mechanical part of the free energy (13) we will formally consider the terms $\boldsymbol{H}^{(s)}\{\mathbf{u}^{(s)}\}$ and $\boldsymbol{H}^{(s,v)}\{\mathbf{u}^{(s)}, \mathbf{u}^{(v)}, w\}$, containing the components of the "static" strain tensor, as a small perturbation with respect to the purely "dynamic" contribution $\boldsymbol{H}_B^{(v)}\{\mathbf{u}^{(v)}, w\}$. Within the first order of perturbation theory, expression (13) reduces to

$$F = F^{(v)} + \boldsymbol{H}^{(s)}\{\mathbf{u}^{(s)}\} + \left\langle \boldsymbol{H}^{(s,v)}\{\mathbf{u}^{(s)}, \mathbf{u}^{(v)}, w\}\right\rangle_{\mathbf{u}, w}.$$

Here, in addition to the quadratic in the "static" strains term (5), two more terms appear: the free energy of purely harmonic oscillations of graphene

$$F^{(v)} = -T \ln \mathrm{Tr}\left(e^{-\boldsymbol{H}_B^{(v)}\{\mathbf{u}^{(v)}, w\}/T}\right) \tag{14}$$

and the average value linear in the components of the "static" strain tensor

$$\left\langle \boldsymbol{H}^{(s,v)}\{\mathbf{u}^{(s)}, \mathbf{u}^{(v)}, w\}\right\rangle_{\mathbf{u}, w} = \frac{\mathrm{Tr}\left(e^{-\boldsymbol{H}_B^{(v)}\{\mathbf{u}^{(v)}, w\}/T} \boldsymbol{H}^{(s,v)}\{\mathbf{u}^{(s)}, \mathbf{u}^{(v)}, w\}\right)}{\mathrm{Tr}\left(e^{-\boldsymbol{H}_B^{(v)}\{\mathbf{u}^{(v)}, w\}/T}\right)}. \tag{15}$$

Note that beyond the perturbation theory the expression for $F$ in (13) by virtue of the Bogolyubov-Peierls inequality is an upper estimate for the true free energy. In any case the values of $\mathbf{u}^{(s)}$ defining the equilibrium thermal strain of graphene are among those delivering a minimum to $F$ in (13).

We emphasize that $\langle \boldsymbol{H}^{(s,v)}\{\mathbf{u}^{(s)},\mathbf{u}^{(v)},w\}\rangle_{\mathbf{u},w}$ determines the contribution to the free energy, which is absent in the quasi-harmonic approximation (for more details, see the next section). Note, that, in addition to $\boldsymbol{H}^{(s)}\{\mathbf{u}^{(s)}\}$, one more quadratic term with respect to the "static" strains could be obtained from the next term of the expansion of (13) over $\boldsymbol{H}^{(s,v)}\{\mathbf{u}^{(s)},\mathbf{u}^{(v)},w\}$. In fact, this term should contribute to the dependence of the elastic moduli of graphene on temperature (this question, as already mentioned, we leave for the future).

Applying the Debye model to a quasi-2D crystal with vibrational in-plane and out-of-plane modes, from (14) we get the following expression for $F^{(v)}$:

$$F^{(v)} = N\varepsilon_0(\rho) + \frac{TS}{\pi}\int_0^{k_{max}} dk\, k \ln\left(1 - e^{-\hbar s_\parallel k/T}\right) + \frac{TS}{2\pi}\int_0^{k_{max}} dk\, k \ln\left(1 - e^{-\hbar \omega_B(k)/T}\right). \tag{16}$$

Here $S$ is the area of the graphene sample, $N$ is the number of unit cells in this area, and $\varepsilon_0(\rho)$ is the energy of zero-point oscillations per unit cell. Since the graphene unit cell contains two atoms we have $\rho = 2mN/S$. Now, along with the "in-plane Debye temperature" (9), let's also define the "out-of-plane Debye temperature":

$$\theta_w \equiv \frac{4\pi\hbar}{m}\sqrt{\kappa\rho}. \tag{17}$$

Substituting into (17) the already given values of the parameters, we obtain $\theta_w = 2040$ K. Using the dimensionless quantities $\xi_\parallel \equiv \theta_\parallel/T$, $\xi_w \equiv \theta_w/T$, and $\xi_B \equiv \hbar s_B^2 \sqrt{\rho/\kappa}/T$, we can reduce expression (16) to the form:

$$F^{(v)} = N\varepsilon_0(\rho) + 2N[2\theta_\parallel f_\parallel(\xi_\parallel) + \theta_w f_w(\xi_w;\xi_B)], \tag{18}$$

where dimensionless functions appear

$$f_\parallel(\xi_\parallel) \equiv \frac{2}{\xi_\parallel^3}\int_0^{\xi_\parallel} d\xi\, \xi \ln\left(1 - e^{-\xi}\right), \quad f_w(\xi_w;\xi_B) \equiv \frac{1}{\xi_w^2}\int_0^{\xi_w} d\xi \ln\left(1 - e^{-\sqrt{\xi^2 + \xi_B\xi}}\right). \tag{19}$$

For the temperatures determined by the inequality $\xi_B \gg 1$, only the interval $0 \leq \xi \lesssim 1/\xi_B$, where the sound dispersion in (11) prevails, contributes to the integral for $f_w(\xi_w;\xi_B)$ in (19). Remembering that $s_B \approx 0.3$ km/s we see that this is the interval $T \ll \hbar s_B^2\sqrt{\rho/\kappa} \sim 1$ K. Hence, at such low temperatures both terms in square brackets of (18) behave as:

$$2\theta_\parallel f_\parallel(\xi_\parallel) + \theta_w f_w(\xi_w;\xi_B) = \frac{m}{4\pi\hbar^2\rho}\left(\frac{2}{s_\parallel^2} + \frac{1}{s_B^2}\right)T^3, \quad T \ll \hbar s_B^2\sqrt{\rho/\kappa}, \tag{20}$$

in agreement with the Nernst theorem for 2D "sound" modes [26]. At the same time, as the numerical calculation shows, the law $T^3$ for the in-plane contribution to (20) is satisfied up to $T \approx 500$ K. It is interesting, however, that for the bending mode there exists a transition region $0.1$ K $\lesssim T \lesssim 40$ K, in which the second term in the dispersion law (11) makes a noticeable contribution to the last term of the free energy (18). This means that only at $T \gtrsim 50$ K one can ignore the contribution of the "sound" part of the spectrum (11) to the thermodynamic functions depending on the bending mode of graphene. In this case, the "low-temperature" dependence $\sim T^2$, which is a consequence of the quadratic dispersion $\omega_B(k) \sim k^2$, will extend to $\approx 500$ K. These results are illustrated in Fig. 1, which presents the effective exponents $d_\| \equiv \partial \ln f_\|(\xi_\|)/\partial \ln T$ and $d_w \equiv \partial \ln f_w(\xi_w;\xi_B)/\partial \ln T$, determining the temperature dependences of the corresponding contributions to the harmonic vibrational free energy of graphene [dashed line – for the effective exponent $d_{w-} \equiv \partial \ln f_w(\xi_w;\xi_B = 0)/\partial \ln T$ built without taking into account the "sound" part of the spectrum (11)].

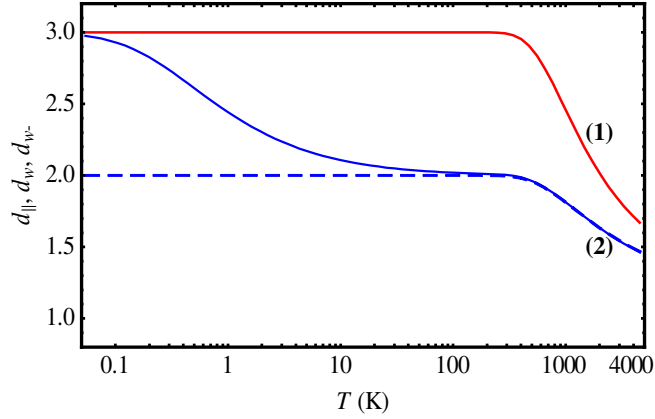

**FIG. 1.** Effective exponents $d_\|$ (curve 1) and $d_w$ (curve 2) determining the temperature dependences of the in-plane and out-of-plane contributions to the harmonic vibrational free energy of graphene. The dashed line shows the effective exponent $d_{w-}$ calculated without taking into account the "sound" part of the out-of-plane spectrum (11).

So, in the most interesting temperature range $T \gtrsim 50$ K, when calculating the function $f_w(\xi_w;\xi_B)$ in (19), one can put $\xi_B = 0$, i.e. $\omega_B(k) = \sqrt{\kappa/\rho}\, k^2$. In this case, the expressions (19) can be rewritten as:

$$f_\|(\xi_\|) = \frac{1}{\xi_\|^3}\left[\xi_\|^2 \ln(1-e^{-\xi_\|}) - \int_0^{\xi_\|} \frac{\xi^2 d\xi}{e^\xi - 1}\right], \quad f_w(\xi_w;\xi_B = 0) = \frac{1}{\xi_w^2}\left[\xi_w \ln(1-e^{-\xi_w}) - \int_0^{\xi_w} \frac{\xi d\xi}{e^\xi - 1}\right].$$

Finally, substituting (6) into (15), averaging the bilinear combinations of the in-plane and out-of-plane strains with respect to the corresponding vibrational modes, and taking into account that, due to the isotropy of graphene, $\langle \partial_\alpha u_\alpha^{(v)} \partial_\beta u_\beta^{(v)}\rangle_\mathbf{u} = \langle \partial_\alpha u_\beta^{(v)} \partial_\beta u_\alpha^{(v)}\rangle_\mathbf{u} = \frac{1}{2}\langle \partial_\alpha u_\beta^{(v)} \partial_\alpha u_\beta^{(v)}\rangle_\mathbf{u}$, we find:

$$\langle \boldsymbol{H}^{(s,v)}\{\mathbf{u}^{(s)},\mathbf{u}^{(v)},w\}\rangle_{\mathbf{u},w} = -\int d\mathbf{r}\, \zeta^{(v)} u_{\alpha\alpha}^{(s)}, \qquad (21)$$

$$\zeta^{(v)} \equiv -\left(\lambda + 2\mu + \frac{3C_{111} + C_{112}}{16}\right)\langle \partial_\alpha u_\beta^{(v)} \partial_\alpha u_\beta^{(v)}\rangle_\mathbf{u} - \frac{\lambda + \mu}{2}\langle (\nabla w)^2\rangle_w. \qquad (22)$$

We recall that the contribution of (21) [with account of (22)] to the free energy of the crystal in the presence of its thermal expansion is absent, in principle, in the quasi-harmonic approximation and, as we will see, it is dominant at "low" temperatures. The quantity $\zeta^{(v)}$, being itself negative, makes a fundamentally *negative* contribution to the GTEC (see below). Moreover, it is clear from the structure of expressions (21) and (22) that they arise due to the explicit presence of *third-order* terms over the strain tensor in the free energy of graphene. To avoid misunderstanding, we emphasize that the resultant sign of $\zeta^{(v)}$ [with the account of inequality (12)] is *negative* precisely due to the anharmonic tensor constructions of the form of $\partial_\alpha u_\beta^{(s)} \partial_\alpha u_\gamma^{(v)} \partial_\beta u_\gamma^{(v)}$ and $\partial_\alpha u_\beta^{(s)} \partial_\alpha w \partial_\beta w$ entering into the "Hamiltonian" (2) with *positive* coefficients $\lambda$ and $\mu$.

At the end of this section, let's give, according to [21], the explicit expression for the average over the out-of-plane oscillations in (22) [the expression for the average over the in-plane oscillations can be restored from (8) and (10)]:

$$\langle (\nabla w)^2 \rangle_w = \frac{\hbar}{2\pi\rho} \int_0^{k_{\max}} \frac{k^3 dk}{\omega_B(k)} \left[ \frac{1}{e^{\hbar\omega_B(k)/T} - 1} + \frac{1}{2} \right]. \tag{23}$$

It is important to recall that just due to the presence of the linear in $q$ term [21] in the dispersion law of the long-wave out-of-plane oscillations of the quasi-2D crystal, the bending mode contribution (23) to the quantity $\zeta^{(v)}$ becomes convergent and thus, ultimately, GTEC is independent of the sample size (see the next Section). At $T \to 0$, in accordance with the Nernst theorem [26], we find from (23):

$$\langle (\nabla w)^2 \rangle_w = \langle (\nabla w)^2 \rangle_{w,T=0} + \frac{\zeta(3)}{\pi\rho\hbar^2 s_B^4} T^3, \tag{24}$$

where $\langle (\nabla w)^2 \rangle_{w,T=0} \approx 0.005$ [21] and $\zeta(3) = 1.202$ is the Riemann zeta-function.

### III. THE THERMAL EXPANSION OF GRAPHENE: BEYOND THE QUASI-HARMONIC APPROXIMATION

Turning to the calculation of contributions into GTEC due to the in-plane and out-of-plane modes, note that the "static" strain tensor in the case of the uniform expansion/contraction has the form $u_{\alpha\beta}^{(s)} = (\delta_{\alpha\beta}/2) u_{\gamma\gamma}^{(s)}$, where $\delta_{\alpha\beta}$ denotes the Kronecker symbol. In this situation, the contributions $\mathbf{H}^{(s)}\{\mathbf{u}^{(s)}\}$ and $\langle \mathbf{H}^{(s,v)}\{\mathbf{u}^{(s)}, \mathbf{u}^{(v)}, w\} \rangle_{\mathbf{u},w}$ [see Eqs. (5) and (21)] are proportional to the "unperturbed" (i.e. at $T = 0$) area $S_0$ of the graphene sample, so:

$$\mathbf{H}^{(s)}\{\mathbf{u}^{(s)}\} + \langle \mathbf{H}^{(s,v)}\{\mathbf{u}^{(s)}, \mathbf{u}^{(v)}, w\} \rangle_{\mathbf{u},w} = S_0 \left( \frac{\lambda + \mu}{2} u_{\alpha\alpha}^{(s)} u_{\beta\beta}^{(s)} - \zeta^{(v)} u_{\alpha\alpha}^{(s)} \right). \tag{25}$$

In the absence of "external" stress, i.e. at free thermal expansion of the graphene sample, the condition of its equilibrium can be written in the form:

$$\frac{\partial}{\partial u_{\alpha\alpha}^{(s)}} \left( F^{(v)} + \mathbf{H}^{(s)}\{\mathbf{u}^{(s)}\} + \langle \mathbf{H}^{(s,v)}\{\mathbf{u}^{(s)}, \mathbf{u}^{(v)}, w\} \rangle_{\mathbf{u},w} \right) = 0. \tag{26}$$

Using the obvious connection between the derivatives (at constant $N$)

$$\frac{\partial F^{(v)}}{\partial u_{\alpha\alpha}^{(s)}} = -\frac{\partial F^{(v)}}{\partial \ln \rho}$$

and expressions (18), (19) and (25), one can obtain from condition (26) the equation determining the temperature dependence of the relative change of the graphene area:

$$u_{\alpha\alpha}^{(s)}(T) = \frac{1}{\lambda+\mu}\left[\frac{\rho^2}{2m}\frac{d\varepsilon_0(\rho)}{d\rho} + \zeta^{(v)}(T) + \frac{2\rho}{m}\left(\theta_\parallel \gamma_\parallel \frac{2}{\xi_\parallel^3}\int_0^{\xi_\parallel}\frac{\xi^2 d\xi}{e^\xi - 1} + \theta_w \gamma_w \frac{1}{\xi_w^2}\int_0^{\xi_w}\frac{\xi d\xi}{e^\xi - 1}\right)\right], \quad (27)$$

where in-plane and out-of-plane Grüneisen parameters are figured

$$\gamma_\parallel \equiv \frac{\partial \ln \theta_\parallel}{\partial \ln \rho}, \quad \gamma_w \equiv \frac{\partial \ln \theta_w}{\partial \ln \rho}. \quad (28)$$

Note that the term $[\rho^2/(2m)]d\varepsilon_0(\rho)/d\rho$ in Eq. (27) is, in fact, determine the "reference point" for the quantity $u_{\alpha\alpha}^{(s)}(T)$; this term, however, does not contribute to GTEC [see Eq. (32) below].

Then from Eq. (27) we find the expression for the GTEC (the subscript $p$ was introduced for agreement with the notation of [25], see below):

$$\alpha_p(T) \equiv \frac{\partial u_{\alpha\alpha}^{(s)}(T)}{\partial T} = \frac{1}{\lambda+\mu}\frac{\partial}{\partial T}\left[\zeta^{(v)}(T) + \frac{2\rho}{m}\left(\theta_\parallel \gamma_\parallel \frac{2}{\xi_\parallel^3}\int_0^{\xi_\parallel}\frac{\xi^2 d\xi}{e^\xi - 1} + \theta_w \gamma_w \frac{1}{\xi_w^2}\int_0^{\xi_w}\frac{\xi d\xi}{e^\xi - 1}\right)\right] \quad (29)$$

and half the value, $\alpha_p(T)/2$, for the linear TEC of graphene. From general considerations it is clear that at high temperatures $\alpha_p(T)$ will tend to some constant value – positive or negative, depending on the relation between the quasi-harmonic, generally speaking, positive second term in square brackets of (29), and the essentially negative value $\zeta^{(v)}(T)$ [see (22)]. To compare both contributions to $\alpha_p(T)$, note that at low temperatures $\zeta^{(v)}(T)$ is determined mainly by the last term in (22), which, with allowance for (24), turns out to be $\sim T^3 s_\parallel^2 / s_B^4$. On the other hand, the main contribution of the quasi-harmonic terms in square brackets of (29) will follow from the term $\sim T^3/s_B^2$ in (20). Thus, taking into account the strong inequality $s_\parallel^2 \gg s_B^2$, we come to the conclusion that the quasi-harmonic contributions to the GTEC at low temperatures are insignificant in comparison with those caused by fundamentally anharmonic [cubic, see (6)] terms in the "Hamiltonian" of graphene.

Based on the results of numerical "experiments" [23], one can estimate the values of the Grüneisen parameters (28) for graphene. Thus, using expression (9) for the "in-plane Debye temperature" $\theta_\parallel$ and introducing the 2D bulk modulus of graphene $B_\parallel \equiv \lambda + \mu$ and the Young modulus $Y \equiv 4\mu(\lambda+\mu)/(\lambda+2\mu)$, we obtain:

$$\theta_\parallel = 4\hbar B_\parallel \sqrt{\frac{2\pi Y}{m(16B_\parallel^2 - Y^2)}}, \quad \gamma_\parallel = \frac{1}{16B_\parallel^2/Y^2 - 1}\left[\frac{1}{2}\left(16\frac{B_\parallel^2}{Y^2} + 1\right)\frac{\partial \ln Y}{\partial \ln \rho} - \frac{\partial \ln B_\parallel}{\partial \ln \rho}\right]. \quad (30)$$

Figs. 3 and 4 of Ref. 23 demonstrate the "experimental" dependences of $B_\parallel(\tau)$ and $Y(\tau)$ on the so-called "mechanical tension $\tau$"; whence we obtain $B_\parallel/Y \approx 0.55$ [23]. Then, taking into account the connection between the derivatives $\partial/\partial \ln \rho \to B_\parallel \partial/\partial \tau$, one can restore from [23] the values $\partial B_\parallel / \partial \tau \approx 6.2$ and $\partial Y / \partial \tau \approx 12.5$. Substituting these values into Eqs. (30) yields the estimate $\gamma_\| \approx 3.7$. This estimate can be reduced taking into account real shapes of the curves for $B_\|(\tau)$ and $Y(\tau)$ shown in Figs. 3 and 4 of Ref. 23.

Similarly, by processing the dependence for the bending rigidity, $\kappa(\tau)$, shown in Fig. 1 of Ref. 23, we can obtain an estimate for the out-of-plane Grüneisen parameter. Representing the latter one with the help of (28) and (17) in the form:

$$\gamma_w = \frac{1}{2}\left(1 + \frac{\partial \ln \kappa}{\partial \ln \rho}\right) \to \frac{1}{2}\left(1 + \frac{B_\|}{\kappa}\frac{\partial \kappa}{\partial \tau}\right), \tag{31}$$

choosing the value of the derivative $(\partial \kappa / \partial \tau) \approx 0.43$ Å$^2$ from the stability region on the curve of $\kappa(\tau)$ (this is the graphene extension region, see Fig.1 of Ref. 23) and substituting the values of other parameters (see above) into (31), we arrive at an acceptable value $\gamma_w \approx 2$.

Turning to the comparison of the found above temperature dependences of the thermal expansion of graphene with the experimental data, we note that the results of measurements of TEC on real samples of graphene published so far are not numerous [8-11] and are largely contradictory. So we shall compare our calculations with the results of numerical "experiments" [22-25]. In this case, the quantities $\gamma_\|$ and $\gamma_w$ will be considered as free parameters, the values of which must be determined from the best agreement between the theory and "experiment". As the latter one, we use the most recent data [25] on the thermal properties of graphene obtained at 12 K $< T <$ 2000 K by the so-called "path-integral molecular dynamics" simulations with simulation cells containing up to 33,600 carbon atoms.

In Fig. 2 we show the found from (27) temperature variation of the area per atom in the projection on the "unperturbed" graphene plane:

$$A_p(T) = A_p(0)[1 + u^{(s)}_{\alpha\alpha}(T) - u^{(s)}_{\alpha\alpha}(0)], \tag{32}$$

where $A_p(0) = 2.6407$ Å$^2$ [25] is the area per atom at $T = 0$ (with the account of zero-point oscillations). The agreement of the theoretical results (solid line) with the "experimental" data for the graphene cell of 33,600 atoms (triangles in Fig. 4 from [25]) was achieved by taking the values of the parameters given above and setting $\gamma_\| = 1.6$, $\gamma_w = 2$ (which seems quite realistic), and $(3C_{111} + C_{112})/8 = -390$ N/m (see above).

An interesting situation arises if we leave only the "quasi-harmonic" term in the square brackets of (27), eliminating the contribution $\zeta^{(v)}(T)$. In this case, instead of (32), we get the expression:

$$A(T) = A(0)\left[1 + \frac{2\rho}{m(\lambda + \mu)}\left(\theta_\| \gamma_\| \frac{2}{\xi_\|^3}\int_0^{\xi_\|}\frac{\xi^2 d\xi}{e^\xi - 1} + \theta_w \gamma_w \frac{1}{\xi_w^2}\int_0^{\xi_w}\frac{\xi d\xi}{e^\xi - 1}\right)\right], \tag{33}$$

where the "reference value" $A(0)$ already contains the contribution of zero-point oscillations and, generally speaking, must not coincide with the value figuring in (32). If we now set $A(0) = 2.6459$ Å$^2$ [25] and represent the dependence (33) in Fig. 2 (the dashed line), then it with good accuracy will reproduce the "experimental" data [25] (empty circles) for the temperature behavior of a "real area $A$ in 3D space" (i.e., the area of the fluctuating surface curved due to the existence of the intrinsic ripples in graphene). Although the foregoing construction is not a strict justification for the fact that in this way we obtain a positive TEC for the "real area" of graphene at all temperatures,[4] the very fact of a close numerical agreement of the results of the quasi-harmonic approximation with molecular dynamics simulations [25] seems worthy of attention.

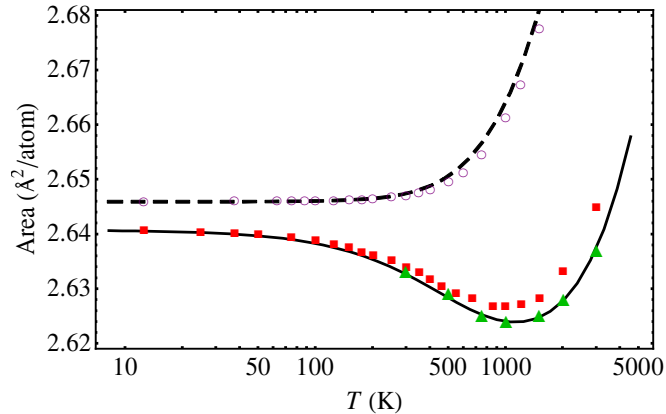

**FIG. 2.** Theoretical [solid line calculated by formulas (32), (27)] temperature dependence of the projected area per graphene atom. Triangles are the data of numerical "experiments" [25] for the graphene sample of 33,600 atoms; squares – the same for the sample of 960 atoms. The experimental [25] temperature dependence of the area per atom of the fluctuating surface curved due to the existence of the intrinsic ripples in graphene is shown by empty circles; the dashed line represents the calculation by Eq. (33).

In Fig. 3 we show the results of calculation of the graphene areal TEC $\alpha_p(T)$ by formula (29) (solid line); triangles designate the values of TEC, restored by numerical differentiation of the "experimental" data [25] for 33,600 atoms (indicated by triangles in Fig. 2). For comparison, filled squares in Fig. 3 show the "experimental" [25] values of $\alpha_p(T)$ for graphene cell including 960 atoms. The role of principle, playing by the non-quasi-harmonic terms in the GTEC at low temperatures is clearly demonstrated by this picture. In addition, the curve $\alpha(T) = \partial \ln A(T)/\partial T$ constructed by formula (33) is also represented in Fig. 3, whereas the empty squares correspond to the "experimental" [25] TEC values for the "real area" of graphene. A rather good coincidence of the results of calculations according to the formula (33) with the "experimental" data [25] deserves attention.

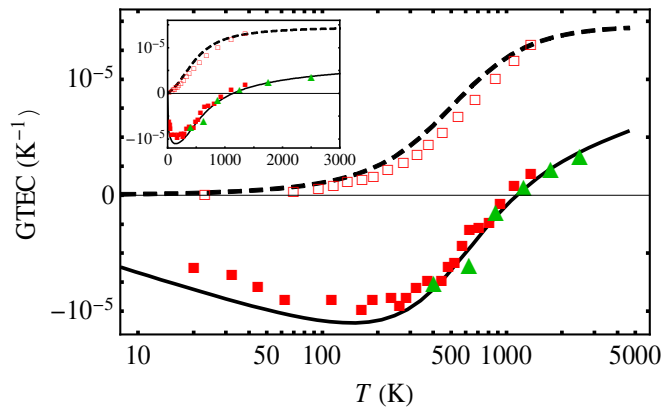

**FIG. 3.** The theoretical (solid line) dependence of the graphene areal TEC $\alpha_p(T)$ calculated by formula (29); the values of GTEC, restored by numerical differentiation of the "experimental"

---

[4] A monotonous increase of the C-C distance in free-standing graphene at temperatures up to 2000 K was obtained in [33] on the basis of the results of *ab initio* molecular dynamics calculations.

data [25] for 33,600 atoms are indicated by triangles. For comparison, filled squares in Fig. 3 show the "experimental" [25] values of $\alpha_p(T)$ for graphene cell including 960 atoms. The "experimental" [25] TEC values for the "real area" of graphene are designated by empty squares. In addition, the curve $\alpha(T) = \partial \ln A(T)/\partial T$ constructed by formula (33) is represented by the dashed line. In the insert, the same results are shown using the linear temperature scale.

## IV. CONCLUSIONS

In this paper the peculiarities of thermal expansion of graphene were investigated on the base of nonlinear theory of elasticity of quasi-2D continuum. A specific feature of the theory of thermal expansion developed here consists in the essential going beyond the framework of quasi-harmonic approximation with explicit inclusion into the Hamiltonian of the elastic medium under consideration the terms of the third order over the strain tensor. As a result, the "cross" terms that connect the static (responsible for the thermal expansion) strains with the mean square fluctuations in the strain fields of vibrational modes arise in the free energy of the quasi-2D medium. A special role is played here by the bending mode, which, as was shown in the authors' paper [21], should have the "sound" dispersion in the long wave region. Due to it, the contribution of the bending mode to GTEC turns out to be *finite* (whereas in models with quadratic dispersion for small wave vectors such a contribution *diverges*). Being definitely negative, this contribution, ultimately, provides the negative sign of GTEC at relatively low temperatures.

The results obtained in the present paper make it possible to clarify the question concerning the meaning of the Grüneisen parameters used when modeling TEC of solids, in particular, of graphene. Usually the description of the thermal expansion of a solid is based on the quasi-harmonic approximation (see Introduction), and, for example, in the framework of the Debye model, the sign of TEC is directly determined by the sign of the Grüneisen constant, i.e. by the derivative of the Debye temperature with respect to density (or pressure) [26]. In fact, this means that in cases where the TEC of a material can change sign with temperature, for a consistent interpretation of the phenomenon of thermal expansion it is required to go beyond the quasi-harmonic approximation. In the present paper it was shown that an explicit account of the anharmonic contributions to the elastic energy of a quasi-2D crystal, together with quasi-harmonic terms, allows a quantitative interpretation of the peculiarities of the thermal expansion of graphene. It is significant that the thermal contraction of graphene in the quantum region of "low" (up to $T \approx 500$ K or even higher) temperatures is associated not with the negative sign of any of the Grüneisen constants (both of them, $\gamma_\parallel$ and $\gamma_w$, are positive, see above) but with the presence in the equation of equilibrium of a fundamentally *negative* contribution due to the anharmonic terms in the "Hamiltonian" of graphene. There is hardly any reason to link this contribution with some *extra* "Grüneisen parameter" having the negative sign. It is more natural to keep the term "Grüneisen parameter" as a characteristic of the quasi-harmonic approximation where the dependence of the thermodynamic quantities of a solid on its volume is, in fact, directly related to a change in its density. Meanwhile, accounting for anharmonic (cubic) terms in the free energy leads to the appearance of explicit terms linear in the static strain tensor of a quasi-2D lattice. In view of the fact that these terms are inherently different from those contained in the quasi-harmonic approximation, there is no longer required to attribute any special temperature dependence to the Grüneisen parameters when such terms are taken into account. In this case, for a consistent theoretical description of the thermal expansion of graphene over a wide temperature range, it is necessary to take into account the above-mentioned anharmonic terms in the free energy, whereas the Grüneisen parameters in the quasi-harmonic approximation should be considered as *positive constants*, the estimate of which can be obtained from the "experimental" data [23].

Note that in the present paper we did not dwell on the question of the possible temperature dependence of the graphene elastic moduli. In principle, such a question could be posed already at the level of quartic terms of the form $\partial_\alpha u_\gamma^{(s)} \partial_\beta u_\gamma^{(s)} \partial_\alpha u_\delta^{(v)} \partial_\beta u_\delta^{(v)}$ [we did not write them out in (6)]; averaging them over the in-plane sound modes would lead to the appearance of temperature dependencies of $\lambda$ and $\mu$ moduli. Meanwhile, similar constructions will appear in the elastic "Hamiltonian" due to the fourth-order terms upon the strain tensor, as well; such terms will also contribute to the dependence of $\lambda$ and $\mu$ on temperature. The account of the above quartic terms could lead to a refinement of the results obtained on the thermal expansion of graphene at high temperatures. On the other hand, a theoretical calculation of the temperature dependences of the elastic moduli of quasi-2D graphene-type lattices could also be of independent interest.

Finally, it seems useful to extend numerical simulations for quasi-macroscopic graphene samples to temperatures below 300 K and compare the results with the predictions of the theory proposed in this paper.

## ACKNOWLEDGMENTS

This work was supported by the Ministry of Education and Science of Ukraine, Grants #0115U003208 and #0117U001111.